\documentclass{elsart}
\usepackage{graphicx}
\usepackage{dcolumn}
\usepackage{bm}

\begin{document}

\begin{frontmatter}

\title{Sigma-omega meson coupling and properties of nuclei and nuclear matter}

\author{Maryam M. Haidari} and \author{Madan M. Sharma\corauthref{cor1}}
\address{Physics Department, Kuwait University, Kuwait 13060}

\ead{sharma@kuc01.kuniv.edu.kw}
\corauth[cor1]{Corresponding author}
\date{\today}

\begin{abstract}

We have constructed a Lagrangian model with a coupling of 
$\sigma$ and $\omega$ mesons  in the relativistic mean-field theory. 
Properties of finite nuclei and nuclear matter are explored 
with the new Lagrangian model SIG-OM. The study shows that an
excellent description of binding energies and charge radii of nuclei over 
a large range of isospin is achieved with SIG-OM. With an incompressibility
of nuclear matter $K=265$ MeV, it is also able to describe the breathing-mode 
isoscalar giant monopole resonance energies appropriately. 
It is shown that the high-density behaviour of the equation of state 
of nuclear and neutron matter with the $\sigma$-$\omega$ coupling 
is much softer than that of the non-linear scalar coupling model.

\end{abstract}


\end{frontmatter}

\maketitle

\section{Introduction}

The relativistic mean-field (RMF) theory \cite{SW.86,Rein.89,Ser.92,Ring.96}
has been a successful approach to describing properties of nuclei along 
the stability line as well as far away from it \cite{GRT.90,SNR.93,Lala.97}. 
Due to the Dirac-Lorentz structure of spin-orbit interaction, it has been 
shown to be advantageous over the conventional non-relativistic Skyrme 
theory in describing properties such as anomalous isotope shifts in Pb
nuclei \cite{SLR.93}. It was shown \cite{SLK.94} that an isospin dependence of 
the spin-orbit interaction or essentially a lack of it is responsible for the 
anomalous behaviour of the isotope shifts. The non-relativistic approaches
based upon the Skyrme and Gogny forces have been unable to reproduce the 
kink \cite{Tajima.93}. On the other hand, by significant alterations in the 
isospin dependence of the spin-orbit potential in the Skyrme theory
\cite{RF.95}, it becomes possible to describe the above anomaly in isotope 
shifts. 

The prevalent model within the RMF theory is that of non-linear couplings of
$\sigma$ meson \cite{BB.77}. Herein, the nonlinear scalar self-couplings of
$\sigma$ meson play a pivotal role in describing finite nuclei. It was
realized earlier on \cite{BB.77} that  $\sigma^3$ + $\sigma^4$ terms
have ingredients appropriate for a proper description of the surface properties
of nuclei. By virtue of this, the scalar self-coupling terms 
have become indispensable for finite nuclei. One of the first nuclear forces
constructed for calculation of ground-state properties of finite nuclei
was NL1 \cite{Rein.89}. A larger value of the asymmetry energy of NL1, 
however, made its application to nuclei far away from the stability 
line difficult.
Within this model, one of the first successful nuclear forces was NL-SH 
\cite{SNR.93} which was obtained with a significant improvement in the
asymmetry energy. The force NL-SH has an incompressibility of nuclear matter
which is on the higher side of physical acceptable region. Sunsequently,
force NL3 was obtained \cite{Lala.97} with an improvement in the nuclear 
incompressibility of nuclear matter. This has in meanwhile become successful 
in several respects. 

A problem that afflicts the model with $\sigma^3$ + $\sigma^4$, in general, is 
that it gives an equation of state (EOS) of nuclear and neutron matter that is 
very stiff. Consequently, it is considered as incompatible with the observed 
spectrum of neutron star masses. A quartic term $\omega^4$ was added recently 
to the above model that was intended to improve the shell effects in nuclei
\cite{Sharma.00}. It has subsequently been shown that the addition of the 
vector self-coupling of $\omega$ meson has the advantage of softening the EOS 
of nuclear matter \cite{Suga.94,Toki.98}. 

Variations in the relativistic approach based upon point-coupling model 
\cite{Nikolaus.92,Rusnak.97,Buerv.02} or of density-dependent meson couplings 
\cite{Brock.92,Typel.99,Typel.05,Niks.02,Long.04} have been made. With the 
inclusion of a larger number of parameters a good degree of success has 
been achieved \cite{Niks.02}. There has also been significant discussion on 
the use of the effective field theoretical approach based upon 
expansion of the interaction Lagrangian in higher order terms in fields 
\cite{Savush.97,Furn.97,Serot.04}. Addition of higher order terms 
and cross terms between the fields are shown to improve description of 
properties of finite nuclei \cite{Furn.97}. Recently, a coupling 
between $\sigma$ and $\rho$ meson has shown to be promising in reduction of 
neutron skin to reasonably accepted values as well as for obtaining a softer 
EOS of nuclear matter \cite{Todd.05}. Thus, within the effective Lagrangian
approach, the field of the RMF theory has become widely open with many
possibilities of interaction terms. It is, however, not known as to what 
form of density dependence of nucleon-meson interactions or of meson-meson
interactions would be the most suitable one. In other words, which terms
nature prefers in the expansion scheme is not known a priori. Emergence 
of nuclear data in extreme regions of the periodic table including those on 
finite nuclei, giant resonances and neutron star masses put nuclear 
interactions to test. 

With the initial success of the RMF theory in nuclear structure, 
there is thus a strong need to construct a theory that should be 
appropriate for broader aspects of finite nuclei as well as for 
nuclear matter. As improved and refined predictions of nuclear 
properties especially the masses (binding energies) of nuclei are 
required for regions often very far away from the stability line
such as those for r-process nucleosynthesis, it is pertinent to 
devise new models and approaches to deliver the same.
This may  include other degrees of freedom or alternative forms of 
interactions. 

In this work, we have explored the RMF Lagrangian in the effective Lagrangian 
approach with a limited number of terms in the expansion with a view to see 
as to whether such an approach is feasible. We have restricted 
ourselves to a meson-meson interaction term between $\sigma$ and 
$\omega$ mesons. Specifically, we have added a coupling of the form 
$\sigma^2 \omega^2$, in addition to the $\sigma^3$ + $\sigma^4$ couplings 
which have so far been the most important terms in the RMF Lagrangian
and are indispensable for finite nuclei and nuclear matter.

\section{Relativistic Mean-Field Theory: Formalism}

The basic (linear) RMF Lagrangian that describes nucleons as Dirac spinors 
interacting with the meson fields is given by \cite{SW.86}
\begin{eqnarray}
{\cal L}_0&=& \bar\psi \left( \rlap{/}p - g_\omega\rlap{/}\omega -
g_\rho\rlap{/}\vec\rho\vec\tau - \frac{1}{2}e(1 - \tau_3)\rlap{\,/}A -
g_\sigma\sigma - M_N\right)\psi\nonumber\\
&&+\frac{1}{2}\partial_\mu\sigma\partial^\mu\sigma- \frac{1}{2}m^2_\sigma \sigma^2
-\frac{1}{4}\Omega_{\mu\nu}\Omega^{\mu\nu}+ \frac{1}{2}
m^2_\omega\omega_\mu\omega^\mu\\ &&
-\frac{1}{4}\vec R_{\mu\nu}\vec R^{\mu\nu}+
\frac{1}{2} m^2_\rho\vec\rho_\mu\vec\rho^\mu -\frac{1}{4}F_{\mu\nu}F^{\mu\nu},
\nonumber
\end{eqnarray}
where $M_N$ is the bare nucleon mass and $\psi$ is its Dirac spinor. 
Nucleons interact with $\sigma$, $\omega$, and $\rho$ mesons, where
$g_\sigma$, $g_\omega$, and $g_\rho$ are the respective coupling constants. 
The photonic field is represented by the electromagnetic vector $A^\mu$. 

The nonlinear $\sigma$-meson-couplings added by Boguta and Bodmer \cite{BB.77}
are of the form
\begin{eqnarray}
U_{NL} = \frac{1}{3}g_2\sigma^3_{} 
+ \frac{1}{4}g_3\sigma^4. 
\end{eqnarray}
The parameters $g_2$ and $g_3$ are the nonlinear couplings of $\sigma$-meson
in the conventional $\sigma^3$ + $\sigma^4$ model. The effective Lagrangian 
that is commonly used is
\begin{eqnarray}
{\cal L}_{eff} = {\cal L}_0  +  U_{NL}
\end{eqnarray}
Here, we add an interaction term between the $\sigma$ and $\omega$
meson. Then, the effective Lagrangian becomes
\begin{eqnarray}
{\cal L}_{eff} = {\cal L}_0  +  U_{NL}+ 
 \frac{1}{2}g_{\sigma\omega} \sigma^2 \omega_{\mu}\omega^{\mu}.  
\end{eqnarray}
The last term represents the coupling between $\sigma$ and $\omega$ meson that 
we introduce. The constant $g_{\sigma\omega}$ is the coupling constant of $\sigma$ and
$\omega$ meson-meson interaction. The field tensors of the vector mesons and 
of the electromagnetic
field take the following form:
\begin{equation}
\begin {array}{rl}
\Omega^{\mu\nu} =& \partial^{\mu}\omega^{\nu}-\partial^{\nu}\omega^{\mu}\\
{\bf R}^{\mu\nu} =& \partial^{\mu}
                  \mbox{\boldmath $\rho$}^{\nu}
                  -\partial^{\nu}
                  \mbox{\boldmath $\rho$}^{\mu}\\
F^{\mu\nu} =& \partial^{\mu}{\bf A}^{\nu}-\partial^{\nu}{\bf A}^{\mu}.\\
\end{array}
\end{equation}
The variational principle gives rise to the Dirac equation:
\begin{equation}
\{ -i{\bf {\alpha}}.\nabla + V({\bf r}) + \beta m*  \}
~\psi_{i} = ~\epsilon_{i} \psi_{i},
\end{equation}
where $V({\bf r})$ represents the $vector$ potential:
\begin{equation}
V({\bf r}) = g_{\omega} \omega_{0}({\bf r}) + g_{\rho}\tau_{3} {\bf {\rho}}
_{0}({\bf r}) + e{1-\tau_{3} \over 2} {A}_{0}({\bf r}),
\end{equation}
and $S({\bf r})$ is the $scalar$ potential
\begin{equation}
S({\bf r}) = g_{\sigma} \sigma({\bf r})
\end{equation}
which defines the effective mass as given by
\begin{equation}
m^{\ast}({\bf r}) = m + S({\bf r}).
\end{equation}
The corresponding Klein-Gordon equations can be written as
\begin{eqnarray}
( -\Delta + m_{\sigma}^{*2} ) \sigma & = & -g_\sigma\bar\psi\psi \\
( -\Delta + m_{\omega}^{*2} ) \omega_\nu & = & g_\omega\bar\psi\gamma_\nu\psi \\
( -\Delta + m_{\rho}^{2} ) \vec\rho_\nu & = & g_{\rho} \bar\psi\gamma_\nu\vec\tau\psi \\
 -\Delta A_\nu & = & \frac{1}{2}e\bar\psi(1 + \tau_3)\gamma_\nu\psi,
\end{eqnarray}
where,
\begin{eqnarray}
m_{\sigma}^{*2} & = & m_{\sigma}^2 + g_2\sigma 
                   + g_3{\sigma}^2 - g_{\sigma\omega}{\omega_0}^2 \\
m_{\omega}^{*2} & = & m_{\omega}^2 + g_{\sigma\omega}\sigma^2.
\end{eqnarray}
These equations imply an implicit density dependence of $\sigma$ and $\omega$
meson masses. These density dependences are then responsible for the density
dependence of nuclear interaction in a nucleus. This is in contrast to the
nonrelativistic Skyrme approach wherein the density functional is well-defined
at the outset. An explicit density dependence of mesons has been introduced
in refs. \cite{Typel.99,Niks.02,Long.04}. This approach requires several
additional parameters besides the usual coupling constants in order to
model a density dependence whose form is not known a priori. 

For the case of an even-even nucleus with a time-reversal symmetry, the
spatial components of the vector fields 
\mbox{\boldmath $\omega$}, \mbox{\boldmath $\rho_3$} and
\mbox{\boldmath A} vanish. The Klein-Gordon equations for the meson 
fields are then time-independent inhomogeneous equations with the 
nucleon densities as sources:
\begin{equation}
\begin{array}{ll}
( -\Delta + m_{\sigma}^{2} )\sigma({\bf r})
 =& -g_{\sigma}\rho_{s}({\bf r})
-g_{2}\sigma^{2}({\bf r})-g_{3}\sigma^{3}({\bf r}) + 
           g_{\sigma\omega}\sigma ({\bf r}) \omega^2_0 ({\bf r}) \\
  (-\Delta + m_{\omega}^{2} ) \omega_{0}({\bf r})
=& g_{\omega}\rho_{v}({\bf r}) - g_{\sigma\omega} \sigma^2({\bf r})  \omega_0 ({\bf r}) \\
    ( -\Delta + m_{\rho}^{2} )\rho_{0}({\bf r})
=& g_{\rho} \rho_{3}({\bf r})\\
\  -\Delta A_{0}({\bf r}) = e\rho_{c}({\bf r}).
\end{array}
\end{equation}
For the mean-field, the nucleon spinors provide the corresponding source terms:
\begin{equation}
\begin{array}{ll}
\rho_{s} =& \sum\limits_{i=1}^{A} \bar\psi_{i}~\psi_{i}\\
\rho_{v} =& \sum\limits_{i=1}^{A} \psi^{+}_{i}~\psi_{i}\\
\rho_{3} =& \sum\limits_{p=1}^{Z}\psi^{+}_{p}~\psi_{p}~-~
\sum\limits_{n=1}^{N} \psi^{+}_{n}~\psi_{n}\\
\rho_{c} =& \sum\limits_{p=1}^{Z} \psi^{+}_{p}~\psi_{p},
\end{array}
\end{equation}
where the sums are taken over the valence nucleons only. 

The stationary state solutions $\psi_i$ are obtained from the coupled 
system of Dirac and Klein-Gordon equations self-consistently.
The solution of the Dirac equation is obtained by using the method of
oscillator expansion \cite{GRT.90}. The ground-state of the nucleus is 
described by a Slater determinant $\vert\Phi>$ of single-particle
spinors $\psi_i$ (i = 1,2,....A). 

The centre-of-mass correction is added using the harmonic oscillator
estimate:
\begin{equation}
E_{c.m.} = \frac{3}{4}.41A^{-1/3}.
\end{equation}

\section{The Lagrangian Parameter set}

The parameters of the new Lagrangian model are obtained by fitting binding
energies and charge radii of a set of nuclei within the RMF theory.
The procedure has been described in ref. \cite{SNR.93}. 
The nuclei included are $^{16}$O, $^{40}$Ca, $^{48}$Ca, $^{90}$Zr, 
$^{116}$Sn, $^{124}$Sn, $^{132}$Sn, and $^{208}$Pb. 
The Sn isotopes $^{116}$Sn, $^{124}$Sn and $^{132}$Sn are included with a 
view to take into account the broad range of isospin dependence of the
nuclear interaction. For open-shell nuclei, the pairing is included 
within the BCS scheme, where the pairing gaps are obtained from the 
experimental masses of neighbouring nuclei. 

The binding energies and charge radii of nuclei have been used as constraints.
In addition, we have tagged the spin-orbit splitting of $p_{3/2}-p_{1/2}$ 
in $^{16}$O. However, we have avoided putting explicit conditions or 
constraints on the nuclear matter properties. The $\omega$ and $\rho$ meson 
masses have been fixed at their empirical values. The procedure of obtaining 
the parameter sets of the new Lagrangian is the same as that applied in 
obtaining other forces such as NL-SH and NL3. It may recalled that the set 
NL3 was obtained by putting constraints explicitly on the nuclear matter 
properties as well.

\noindent
\begin{table}[h*]
\begin{center}
\caption{The parameters of the Lagrangian SIG-OM with the coupling between 
$\sigma$ and $\omega$ mesons. The Lagrangian sets with the non-linear 
scalar self-coupling NL-SH, NL3 are shown for comparison.}
\bigskip
\begin{tabular}{l l l l l }
\hline
& Parameters~~~        & SIG-OM~~~~~      & NL-SH~~~~~     &  NL3~~~~~ \\    
\hline 
&M (MeV)            & 939.0       & 939.0     & 939.0    \\
&$m_{\sigma}$ (MeV) & 505.9263    & 526.0592  & 508.1941 \\
&$m_{\omega}$ (MeV) & 783.0       & 783.0     & 782.501  \\
&$m_{\rho}$ (MeV)   & 763.0       & 763.0     & 763.0    \\
&$g_{\sigma}$       & 10.0429     & 10.4436   &  10.2169 \\
&$g_{\omega}$       & 12.7668     & 12.9451   &  12.8675 \\
&$g_{\rho}$         & 4.4752      & 4.3828    &   4.4744 \\
&$g_{2}$ (fm$^{-1}$)& $-$7.9223   & $-$6.9099  & $-$10.4307 \\
&$g_{3}$            & 12.4601     & $-$15.8337 & $-$28.8851 \\
&$g_{\sigma\omega}$ & 35.6922     & ~~0.0      &  ~~0.0   \\ 
\hline
\end{tabular}
\end{center}
\end{table}
The parameters of the Lagrangian thus obtained (here named as SIG-OM) 
are listed in Table 1. These are compared with those of the forces 
NL-SH and NL3 with
the nonlinear scalar self-couplings. The coupling constant $g_3$ with SIG-OM
is positive in contrast with that of NL-SH and NL3, where it is predominantly 
negative. It has been argued \cite{Rein.89} that a negative $g_3$ has a 
consequence in that the spectrum of the full theory is not bound 
from below and that renormalization of the scalar field is not possible.  

Solving the equations for nuclear matter at the saturation point, equilibrium
properties of the nuclear matter are calculated. Nuclear properties arising 
from the parameters of SIG-OM are shown in Table 2. A comparison is made with
the properties of NL-SH and NL3. The saturation density and binding 
energy per nucleon for SIG-OM are very close to those of NL-SH and NL3 and 
are in physically acceptable region. The incompressibility of nuclear matter 
$K$ that arises from SIG-OM is 265.2 MeV. It is slightly smaller than that 
of NL3 ($K = 271.6$ MeV). The effective mass at the saturation point with
SIG-OM is $m* =0.62$ . It is slightly bigger than a value of $\sim$0.60
for NL-SH and NL3. The asymmetry energy $a_4$ of SIG-OM is 37.0 MeV.
It is comparable to that of NL-SH and NL3 and is, however, still bigger 
than the received empirical value of $\sim$33 MeV. 

\noindent
\begin{table}
\begin{center}
\caption{The nuclear matter (NM) properties due to the force SIG-OM. 
Properties of the sets NL-SH and NL3 are also given for comparison.}
\bigskip
\begin{tabular}{l l c c c }
\hline
& NM Properties  & ~~~~SIG-OM~~~~   & ~~~~NL-SH~~~~ &  ~~~~NL3~~~~   \\ 
\hline
& $\rho_0$ (fm$^{-3}$) & 0.149    & 0.146      & 0.148    \\
& $a_v$ (MeV)          & $-$16.30 & $-$16.33   & $-$16.24 \\
 & $K$ (MeV)           & 265.2    & 354.9      & 271.6    \\
& $m^*$                & 0.622    & 0.597      & 0.595    \\
& $a_4$ (MeV)          & 37.0     & 36.1       & 37.4     \\
\hline
\end{tabular}
\end{center}
\end{table}
\section{Results and discussion}

\subsection{Ground-state properties of spherical nuclei}

The binding energies and charge radii of nuclei calculated with SIG-OM 
in the RMF theory are presented in Tables 3. Results with NL-SH and NL3 are 
also given for comparison. Table 3 shows the binding energies 
achieved for the key 
nuclei included in the fit along with those of several other isotopes of 
Sn and Pb. The experimental binding energies \cite{Audi.95} and charge
radii where available are shown in the last column. The binding energies of 
the nuclei included in the fit are reproduced well by SIG-OM with a slight 
overbinding for $^{16}$O. The set includes three doubly magic nuclei of 
$^{100}$Sn, $^{132}$Sn and $^{208}$Pb.
It is important that any newly developed Lagrangian model should be able to
describe the binding energies of closed-shell nuclei. Comparing the binding
energy of $^{132}$Sn and $^{208}$Pb due to SIG-OM with the experimental data,
one can see that there is a good agreement for $^{132}$Sn. For $^{208}$Pb,
SIG-OM overestimates the energy by $\sim 1.5$ MeV. This is, however, much
improved as compared to that of NL-SH and NL3, where these sets overestimate
the binding energy of $^{208}$Pb by $\sim 3-4$ MeV. The binding energies of 
other isotopes of Sn and Pb nuclei obtained with SIG-OM show an excellent 
agreement with the experimental data as compared to NL-SH and NL3. 

The case of $^{100}$Sn needs a special mention. Almost all Lagrangian sets in
use in the RMF theory overestimate the binding energy of $^{100}$Sn
significantly. This includes both NL-SH and NL3 which overestimate the binding 
energy by $\sim 4-5$ MeV. The prediction of SIG-OM for $^{100}$Sn is
surprisingly close to the experimental value. This has, however, not affected 
(underpushed) the binding energies of $^{116}$Sn, $^{124}$Sn and $^{132}$Sn in 
the isotopic chain of Sn. Thus, SIG-OM provides the best available description
of doubly magic nucleus $^{100}$Sn. 
\noindent
\begin{table}
\begin{center}
\caption{The binding energies in MeV and charge radii (parentheses) in fm 
of spherical nuclei obtained with SIG-OM. The values for NL3 and NL-SH are
also shown for comparison. The empirical values (exp.) are shown in the 
last column}
\bigskip
\begin{tabular}{c l l l l l}
\hline
& Nucleus  & ~~~~SIG-OM & ~~~~~NL-SH &  ~~~~~~~NL3 &  ~~~~~~~exp.~~~~   \\    
\hline 
& $^{16}$O   & $-$129.2 (2.699) & $-$128.4 (2.699) & $-$128.8 (2.728) & $-$127.6 (2.730)\\
& $^{40}$Ca  & $-$343.3 (3.440) & $-$340.1 (3.452) & $-$342.0 (3.469) & $-$342.1 (3.450)\\
& $^{48}$Ca  & $-$414.6 (3.454) & $-$415.0 (3.462) & $-$415.1 (3.471) & $-$416.0 (3.451) \\
& $^{76}$Ni  & $-$631.8 (3.920) & $-$634.4 (3.920) & $-$634.1 (3.926) & $-$633.1 \\
& $^{90}$Zr  & $-$783.1 (4.282) & $-$782.9 (4.282) & $-$782.6 (4.287) & $-$783.9 (4.258)\\
& $^{100}$Sn  & $-$826.0 (4.465) & $-$830.6 (4.467) & $-$829.2 (4.473) & $-$824.5 \\  
& $^{116}$Sn  & $-$988.8 (4.593) & $-$987.9 (4.599) & $-$987.7 (4.611) & $-$988.7 (4.626)\\ 
& $^{124}$Sn  & $-$1049.9 (4.645) & $-$1050.1 (4.651) & $-$1050.2 (4.661) & $-$1050.0 (4.673)\\  
& $^{132}$Sn  & $-$1102.9 (4.698) & $-$1105.9 (4.702) & $-$1105.4 (4.708) & $-$1102.9 \\  
& $^{202}$Pb  & $-$1591.8 (5.480) & $-$1596.0 (5.480) & $-$1592.6 (5.497) & $-$1592.2 (5.473)\\
& $^{208}$Pb  & $-$1638.3 (5.506)  & $-$1640.4 (5.509) & $-$1639.6 (5.523) & $-$1636.7 (5.503)\\
& $^{214}$Pb  & $-$1663.8 (5.563) & $-$1664.3 (5.562) & $-$1661.6 (5.581) & $-$1663.3 (5.558)\\
\hline
\end{tabular}
\end{center}
\end{table}

The charge radii of nuclei calculated with SIG-OM are given in parentheses in
Table 3. The SIG-OM values describes the experimental charge radii well. 
This is especially the case for Pb isotopes where SIG-OM values are closer to 
the experimental data. In comparison, NL3 overestimates the experimental
charge radii of Pb isotopes. For the Sn isotopes, NL3 values are, however, 
closer to the experimental data. In comparison, charge radii for heavier
isotopes with NL-SH are in close proximity to the experimental data. 
With the new Lagrangian model SIG-OM, there is a general agreement of its 
charge radii with the experimental data especially for heavier nuclei.

\subsection{The isotopic chain of Sn nuclei}

With the new Lagrangian model SIG-OM, we have performed a case study. 
The isotopic chain of Sn nuclei with the experimental masses available 
from the proton drip-line doubly-magic nucleus $^{100}$Sn to another 
doubly-magic nucleus $^{132}$Sn offers unique data for a complete 
coverage of the shell from $N=50$ to $N=82$. We have performed the 
Relativistic Hartree-Bogoliubov calculations for Sn isotopes. The details
of the method are provided in ref. \cite{Sharma.00}. The Gogny force
D1S has been used in the pairing channel. It has been shown
that the Gogny force is able to represent the pairing properties of 
Sn, Pb and other nuclei successfully \cite{Berger.84}. 

The binding energies of Sn nuclei obtained with SIG-OM are shown in Fig. 1. 
A comparison is made with the predictions of NL3. Here, we have also included 
the results due to NL-SV1 \cite{Sharma.00}. The force NL-SV1 is based upon the 
model with the quartic coupling of $\omega$ meson. The results show a 
significant difference between the predictions of SIG-OM and both the 
other Lagrangian models in the regions near the closed shells. 
Whereas NL-SV1 shows a good agreement with the data in the region 
$106 < A < 132$, NL3 shows a reasonably good agreement with the 
data in the region $106 < A < 126$. Both NL3 and NL-SV1 show significant 
deviations from the experimental values below $A < 106$. For the 
double-magic nucleus $^{100}$Sn, the overbinding amounts to $\sim 4-5$ MeV. 
This problem of overbinding of magic nuclei and by implication the presence 
of a strong shell energy is not limited to theoretical models alone. 
The phenomenon of arches at the magic numbers persists both in microscopic 
theories and mass formulae alike. 

Viewing the binding energy curves in Fig. 1, one can see clearly that 
SIG-OM shows an excellent agreement with the experimental data all over 
the range of the shell. It shows a much smaller deviation from the data 
at $A=100$. The well-known arch-like pattern that is exhibited by 
NL-SV1 and NL3 is reduced significantly with SIG-OM. Thus, SIG-OM 
provides a very good description of the binding energies of the Sn 
isotopes over the whole region between the two magic numbers.
\begin{figure}
\vspace{1.0cm}
\hspace{0.5cm}
\resizebox{0.85\textwidth}{!}{%
  \includegraphics{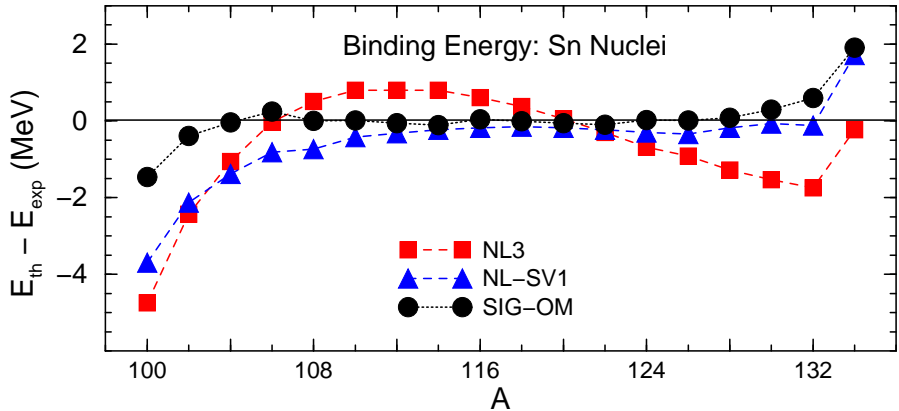}}
\vspace{0cm}       
\caption{The binding energies of Sn isotopes calculated using the
Relativistic Hartree-Bogoliubov method with the Lagrangian SIG-OM. 
The results obtained with NL3 and NL-SV1 are also shown for comparison.}
\label{fig:1}       
\end{figure}

\subsection{Charge radii and isotopic shifts of Pb nuclei}

The salient feature of the Pb chain is the presence of a characteristic 
kink in charge radii and isotopic shifts at the magic number $N=82$. 
This behaviour of charge radii was considered for long to be anomalous 
\cite{Tajima.93}. Thus, the charge radii and isotopic shifts for Pb chain 
have become a standard test bench for any new model. 
In order to view as what response the new model
SIG-OM gives for charge radii, we have performed RMF+BCS calculations 
for the isotopes of Pb. The charge radii of Pb isotopes calculated with 
SIG-OM are shown in Fig. 2. A comparison is made with the  charge radii 
obtained with NL-SH and NL3. The experimental values \cite{Otten.89} are 
shown for a comparison. A kink across $A=208$ ($N=82$) arises in all 
the curves. The SIG-OM values describe the experimental data as well 
as NL-SH does. Whereas both SIG-OM and NL-SH show some 
minor differences with the data for isotopes lighter than 
$^{208}$Pb, for the heavier isotopes SIG-OM values show a better
agreement with the data. The data point for $^{208}$Pb is significant
as charge radius of this nucleus has been measured using various approaches
and its value  ($r_c = 5.503$ fm) is known with a very high precision. 
The charge radius of $^{208}$Pb with SIG-OM is 5.506 fm. It is very close
to the experimental value. The value with NL-SH is 5.509 fm which is slightly
larger than the experimental data. In comparison, the charge radius of 
$^{208}$Pb with NL3 is 5.523 fm. The results of Fig. 2 show that NL3 
overestimates the charge radii of the Pb isotopes systematically. 

\begin{figure}[h*]
\vspace {0.5 cm}
\hspace{0.5cm}
\resizebox{0.90\textwidth}{!}{%
   \includegraphics{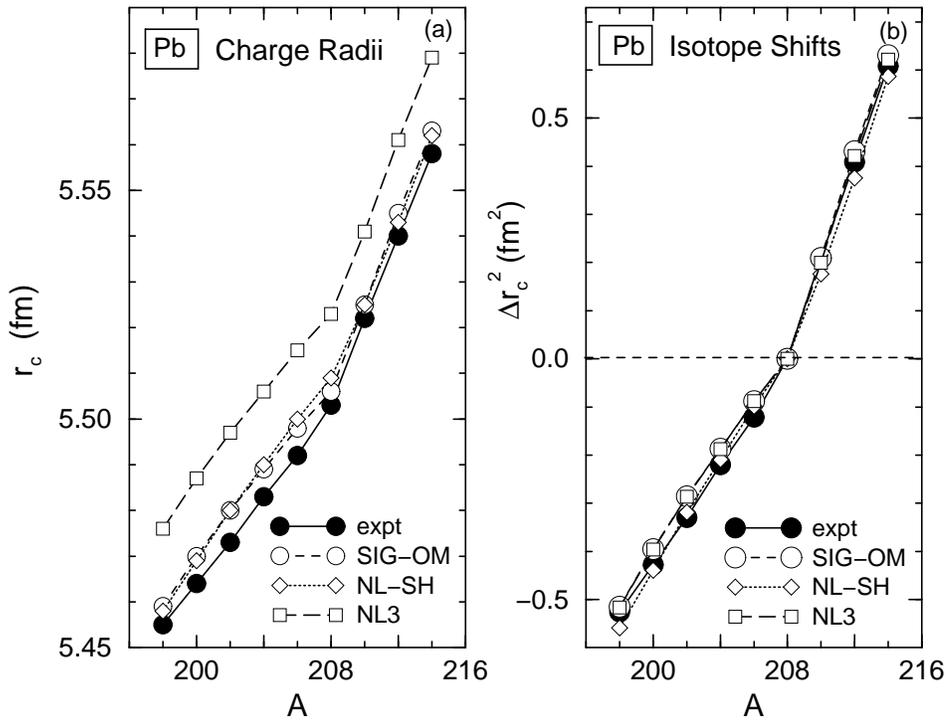}}
\vspace{0cm}       
\caption{The charge radii calculated with SIG-OM, NL-SH and NL3. Experimental
values are also shown for comparison}
\label{fig:2}       
\end{figure}

The isotope shifts $\Delta r_c^2 = r_c^2 (A) - r_c^2(208)$ of Pb nuclei 
have been measured with a significant precision using laser-beam atomic
spectroscopy\cite{Otten.89}. Whilst $^{208}$Pb acts as 
a reference point in the measurement of $\Delta r_c^2$, the actual value 
of the charge radius of $^{208}$Pb does enter into the values of the 
isotopes shifts obtained from the measurements.
We show in Fig. 2(b) the isotopes shifts calculated using the charge radii
from SIG-OM, NL-SH and NL3. The experimental data \cite{Otten.89} are 
shown directly. The theoretical values match the experimental data very 
well irrespective of the Lagrangian model used. Even the larger charge 
radii predicted by NL3 cancel out in the calculation of $\Delta r_c^2$.
Apparently, the isotope shifts of SIG-OM are very close to those due
to NL3. We can conclude that the RMF theory describes the experimental isotope 
shifts very well irrespective of the nature of the Lagrangian model 
employed with barely perceptible differences amongst various forces.

\subsection{Nuclei away from the stability line}

Mass measurements with high precision on nuclei far away from the stability 
are being performed at various facilities and new experimental data are
emerging currently. Such data provide a crucial test bench for various 
nuclear models in regions away from the stability line. In order to test 
the predictive power of the Lagrangian SIG-OM, we have performed 
deformed RMF calculations for some nuclei on which experimental data has
been obtained recently. The total binding energy of some isotopes of 
Si, Sr, and Mo and a few others obtained with SIG-OM is shown in Table 5. 
A comparison is made with NL3. The experimental binding energies obtained 
from high-precision experimental masses on Si 
\cite{Jurado.07}, Sr \cite{Sikler.05}) and Mo \cite{Hager.06}) are 
used for comparison. 

\noindent
\begin{table}[h*]
\begin{center}
\caption{The binding energy E (in MeV) and quadrupole deformation $\beta_2$ 
of nuclei away from the stability line calculated with SIG-OM. 
The deviation of the binding energy from the experimental value, $\Delta$E =
E - E$_{exp}$, is also shown. A comparison is made with the predictions
of NL3.}
\bigskip
\begin{tabular}{l c c c c c c c c}
\hline
& Nucleus     & &SIG-OM & & & & NL3 &     \\    
&             & E  &$\beta_2$   & $\Delta E$ &&  E  & $\beta_2$ & $\Delta E$ \\
\hline 
& $^{36}$Si      & $-$292.4 & 0.02  & -0.4  &&  $-$293.5 & 0.0  & -1.5 \\
& $^{38}$Si      & $-$300.0 & 0.28  & -0.2  &&  $-$301.3 & 0.27 & -1.5 \\
& $^{40}$Si      & $-$306.6 & 0.36  & -0.1  &&  $-$307.9 & 0.34 & -1.4 \\
& $^{42}$Si      & $-$313.3 & -0.35 & -0.4  &&  $-$315.2 & -0.33& -2.3 \\
& $^{80}$Sr      & $-$684.8 & 0.0   & +1.5  &&  $-$682.8 & 0.0  & +3.5 \\
& $^{86}$Sr      & $-$748.9 & 0.0   & 0.0   &&  $-$748.2 & 0.0  & +0.7 \\
& $^{88}$Sr      & $-$768.1 & 0.0   & +0.4  &&  $-$768.0 & 0.0  & +0.5 \\
& $^{108}$Mo     & $-$909.2 & -0.22 & +0.4  &&  $-$908.3 & -0.23 & +1.3 \\
& $^{110}$Mo     & $-$919.6 & -0.23 & -0.1  &&  $-$919.0 & -0.23 & +0.5 \\
& $^{120}$Xe     & $-$1008.2 & 0.26 & +0.3  &&  $-$1007.7 &  0.29 & +0.8 \\
& $^{174}$Yb     & $-$1407.2 & 0.31 & -0.6  &&  $-$1407.2 &  0.32 & -0.6 \\
\hline
\end{tabular}
\end{center}
\end{table}

It is noted that SIG-OM describes the experimental binding energies very 
well with only a few exceptions. The agreement is especially good for 
newly measured Si isotopes. Not surprisingly the force NL3
also shows reasonably good agreement for several data points. It, however,
shows somewhat larger deviations from the data notably for the Si isotopes. 
Details of a comprehensive study to examine the shell structure and shell
effects due to the Lagrangian model SIG-OM encompassing a larger number 
of data in the regions far away from the stability line will be discussed 
elsewhere \cite{Sharma.07}.

\subsection{Breathing mode GMR energies and GCM calculations}

The incompressibility of nuclear matter $K$ is an important point on
the equation of state (EOS) of nuclear matter. By definition, it is the
curvature of the EOS at the saturation point. The value of $K$ is important
as it affects the breathing mode giant monopole resonance (GMR) energies
sensitively. As the force NL-SH has a rather large value, it is found
to be not suitable for the GMR energies of nuclei. It may be recalled that 
within the $\sigma^3$ + $\sigma^4$ model, the force NL3 was obtained with 
a value of incompressibility $K= 271$ MeV which is in the range of physically 
acceptable values. In order to get a comparative picture of the response of
various Lagrangians on the breathing-mode giant monopole resonance (GMR) 
energies, we have performed constrained generator coordinate method (GCM) 
calculations for the GMR mode \cite{Sto.94}. The nuclei included are 
$^{90}$Zr, $^{120}$Sn and $^{208}$Pb. Experimental GMR energies on these 
nuclei are known with a reasonable precision \cite{Young04,Sharma.88}. 
The results of our calculations are shown in Table 6 and are compared 
with the experimental data. Clearly, the force NL-SH ($K=355$ MeV) provides 
larger GMR energies due to its high incompressibility. 
NL3 ($K=271$ MeV), on the other hand, underestimates the GMR data by
$\sim 1$ MeV. Results of our calculations obtained with NL3 agree well with 
those provided in ref. \cite{Lala.97}.

\noindent
\begin{table}[h*]
\begin{center}
\caption{The breathing mode GMR energies in nuclei obtained with the
the constrained GCM calculations. The experimental data are from 
refs. \cite{Young04,Sharma.88}}

\bigskip
\begin{tabular}{c c c c c c c c}
\hline
&   Nucleus    && SIG-OM  & NL3     &  NL-SH      &&  exp. \\    
\hline 
& $^{90}$Zr   && 18.2  & 16.9 & 19.5 && $17.81 \pm 0.30$ \\
& $^{120}$Sn  && 16.2  & 15.0 & 16.7 && $15.52 \pm 0.15$ \\
& $^{208}$Pb  && 14.1  & 13.0 & 15.0 && $13.96 \pm 0.28$ \\
\hline
\end{tabular}
\end{center}
\end{table}

It is interesting to note that SIG-OM with its slightly lower value of 
$K= 265$ MeV than that of NL3 provides GMR energies which are closer to 
the experimental data, especially those of $^{90}$Zr and $^{208}$Pb 
considered frequently in analyses. In comparison, NL3 with a higher 
value of $K=271$ MeV underestimates the data by $\sim $1 MeV. Generally, 
a force with a lower value of $K$ would yield smaller values of GMR 
energies. This difference in the predictions of SIG-OM and NL3 may be
due to a difference in some finite size effect of a nucleus. The other
two factors which play a role for GMR energies are the Coulomb term
and the asymmetry component. At present, it is difficult to estimate
as to which factor brings about the apparent paradoxical behaviour
of SIG-OM and NL3. We believe that it may be due to a different
surface component (arising out of a different density dependence of
the interactions) in SIG-OM vis-a-vis NL3. It will, therefore, be important 
to carry out calculations of surface incompressibility using the 
semi-infinite nuclear matter. These calculations are in progress.
Moreover, it is also to be seen as to what response the two 
Lagrangian models provide in relativistic radom-phase approximation 
calculations of the breathing-mode GMR energies.

\subsection{Nuclear matter properties}

\subsubsection{Effective mass}

The density dependence of the effective mass $m^*$ is to an extent a reflection
of the density dependence of interactions in a nucleus, though it corresponds
to that of the $\sigma$ field to be exact. It would be instructive to see as to
how $m^*$ changes with density in different Lagrangian models. We show in 
Fig. 3 the density dependence of  $m^*$ for different Lagrangian models.
Fig. 3(a) shows the effective mass $m^*$ for the density region of a nucleus
up until a 4-5 times the saturation density. Both NL-SH and NL3 bunch out
together and show a small difference in the high density region. NL-SV1 splits 
out from NL-SH and NL3 at higher densities.  In the high-density region,
three forces NL-SH and NL3 with the scalar self-couplings and NL-SV1 with
the scalar self-coupling along with the vector self-coupling exhibit a
similar pattern of variation with density as seen in Fig. 3(a). 
The model SIG-OM with $\sigma-\omega$ coupling, on the other hand, shows a 
density dependence that is different from that of NL-SH, NL3 and NL-SV1 and
it is especially so in the high density region. 
\begin{figure}
\hspace{0.5cm}
\resizebox{0.90\textwidth}{!}{%
   \includegraphics{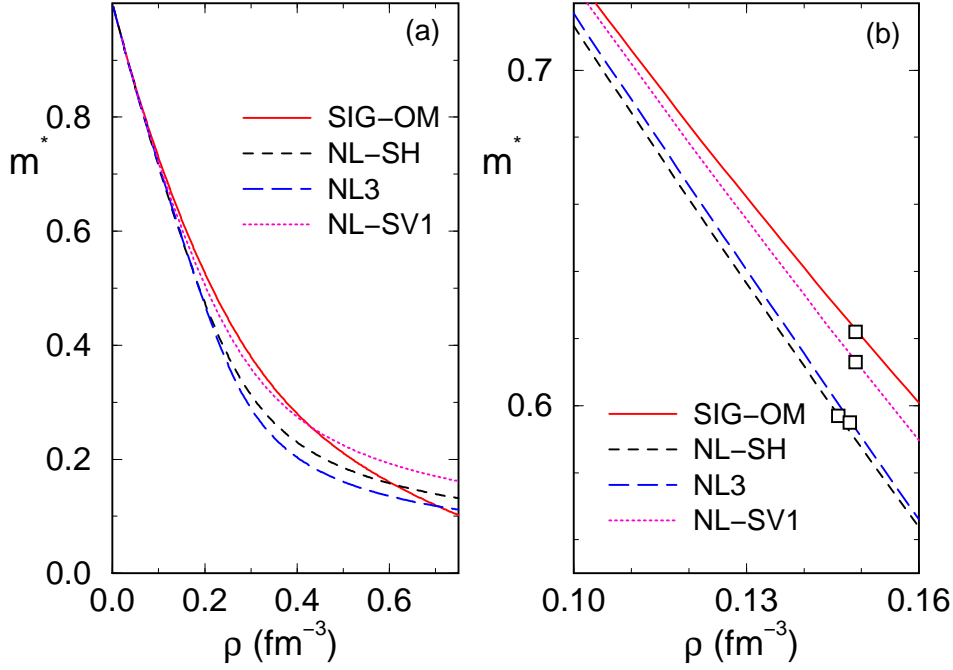}}
\vspace{0cm}       
\caption{The effective mass $m^*$ calculated with SIG-OM. It is compared with
that for the $\sigma^3 + \sigma^4$ Lagrangians NL-SH and NL3. A comparison
is also made with that of $\omega^4$ Lagrangian NL-SV1. (a) The density
dependence of $m^*$ for the whole density density including the high-density
region. (b) $m^*$ in the region of density relevant to finite nuclei.
The saturation point for each case is shown by squares.}
\label{fig:3}       
\end{figure}

The region of the density relevant to finite nuclei is shown in Fig. 3(b).
The saturation point is indicated by a square in each case. Both SIG-OM and
NL-SV1 have a slightly larger $m^*$  than that of NL-SH and NL3. It may be
pointed out though that in each case the spin orbit splitting of $^{16}$O
is reproduced well and yet overall fits to finite nuclei lead to slightly
different effective mass for each model. This is no coincidence that within
the $\sigma^3 + \sigma^4$ Lagrangian model,  the effective masses for NL-SH
and NL3 are very close to each other. The variation in $m^*$ for these
two forces in this region is also very similar. On the other hand, in the 
region of interest for finite nuclei, it is seen clearly that SIG-OM 
shows a density dependence that deviates strongly from that of NL-SH and 
NL3. This may possibly lead to a different features of surface with SIG-OM. 
Detailed surface properties using various Lagrangian models are in progress.
It is a matter for further investigation as to whether this difference 
in density dependence leads to different predictions of breathing-mode
energies with SIG-OM as discussed above.

\subsubsection{ The EOS of nuclear and neutron matter}

The EOS of nuclear and neutron matter is important for structure 
and properties of neutron stars. There has recently been a significant
discussion on EOS of nuclear matter \cite{Klaehn.06} especially in view of 
recent observations of neutron stars which have been found to exist with 
masses in the vicinity of two solar masses or even more \cite{Oezel.06}.
In accordance with the known spectrum of neutron star masses which
lie below $\sim$ 1.8 solar mass ($M_\odot$), the conventional wisdom 
has been that a softer EOS is required for describing the observed 
masses. However, recent discoveries of masses close to $2M_\odot$  
or more than $2M_\odot$ seem to put new demands on the high-density 
EOS of nuclear matter. In ref. \cite{Oezel.06}, the neutron star 
EXO 0748-676 with mass $M \ge 2.10 \pm 0.28 M_\odot$ has been reported. 
Consequently, very  soft EOS for nuclear and neutron matter are ruled out. 
Then, one requires an EOS that should encompass even the heavier neutron 
star masses in the region of $2.0 - 2.3 M_\odot$. 

Solving the equations for nuclear matter self-consistently, we have 
calculated the EOS of nuclear and neutron matter with the new Lagrangian
model. The EOS of nuclear matter with SIG-OM shows a significant 
softening at higher densities 
as compared to that of NL-SH and NL3 (see Fig. 4). The coupling of 
$\sigma$ and $\omega$ meson does have a softening effect at higher densities. 
In contrast, EOS with  the $\sigma^3$ + $\sigma^4$ model is known to be stiff  
due to preponderance of $\omega$ term at higher densities. 

%
\begin{figure}[h*]
\vspace{0.5cm}
\hspace{1.5cm}
\resizebox{0.70\textwidth}{!}{%
      {\includegraphics{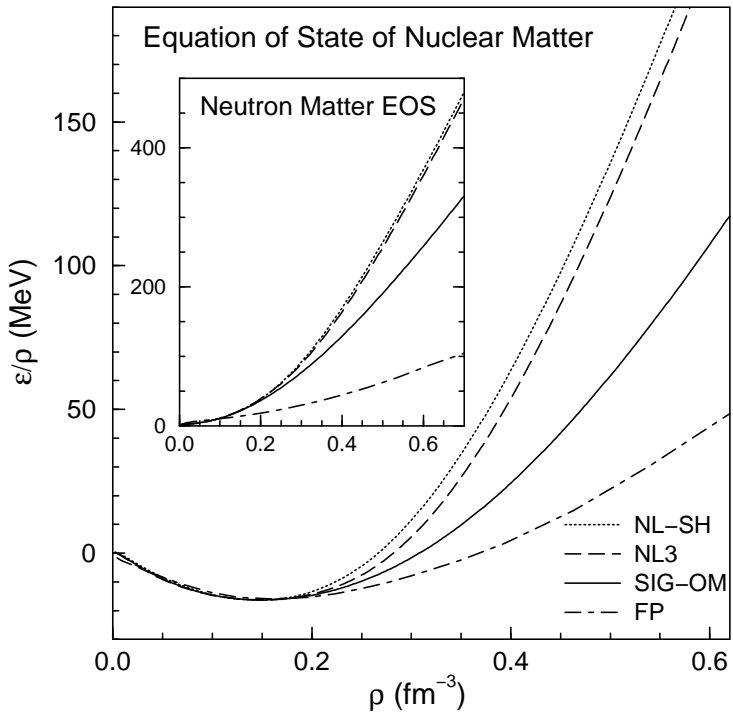}}
}
\vspace{0cm}       
\caption{The EOS of nuclear matter and neutron matter
(inset) obtained with SIG-OM, NL-SH and NL3. A comparison is made with
the EOS due to Friedman and Pandharipande (FP) \cite{FP.81}.}
\label{fig:4}       
\end{figure}

The corresponding EOS for neutron matter is shown in the inset of Fig. 4. 
It can be seen that
neutron EOS due to SIG-OM is also softer than that for NL-SH and NL3. 
For a comparison, we show the EOS due to Friedman and Padharipande 
(FP) \cite{FP.81}. The FP EOS for nuclear as well as neutron matter
is very soft and was intended to describe lower mass neutron stars
with masses $M < 1.6M\odot$. As  one can see from Fig. 4, the softening 
achieved by the $\sigma-\omega$ coupling is not strong enough to match 
that of FP which is a very soft EOS leading to smaller values of maximum 
mass of neutron stars \cite{FP.81}. In view of new observations such as
those of ref. \cite{Oezel.06}, a softer EOS in the high density region may 
not be required. It has to be seen from neutron-star structure
calculations whether the EOS due to SIG-OM which is neither as stiff as
NL3 nor as soft as FP would be suitable for describing the newly observed
neutron star masses.

\section{Conclusions and outlook}

The Lagrangian SIG-OM with the coupling of  $\sigma$ and $\omega$ mesons 
in the RMF theory has been developed. We have explored the feasibility of
adding the $\sigma$-$\omega$ coupling in the RMF Lagrangian. Properties
of finite nuclei and nuclear matter have been explored with the model
SIG-OM. It is shown that SIG-OM provides a very good  description of 
the ground-state properties such as binding energies and charge radii of
nuclei over a large range of isospin. Taking the case of the isotopic chain
of Sn nuclei, it is shown that it describes the binding energies of nuclei
between the two magic numbers $N=50$ and $N=82$ very well. It will be
interesting to see in further investigations as to what implications this 
improvement in shell structural aspects may have in the other regions and 
most importantly in the regions close to the r-process path.

The breathing-mode giant monopole resonance energies for a few nuclei have
been calculated using the constrained GCM calculations. It is shown that
the model SIG-OM with the incompressibility of nuclear matter $K = 265$ MeV 
describes the breathing-mode GMR energies well. A comparative 
analysis of breathing-mode GMR energies with the GCM approach
in the RMF theory suggests that some finite-size effects with SIG-OM may be
different than with Lagrangians with nonlinear scalar self-couplings.
It is important to carry out calculations with relativistic RPA 
approach to see as to whether such a feature would persist. This is a 
matter of further investigations and requires a detailed study.

The equation of state of nuclear and neutron matter have been obtained with
SIG-OM. A comparison is made with the EOS obtained with the interactions
within the nonlinear scalar self-coupling. As it is well-known, the latter
model gives very stiff EOS. The equation of state of nuclear and neutron
matter obtained with the new Lagrangian SIG-OM is significantly softer than 
that with non-linear scalar self-couplings.

We thank Prof. Lev Savushkin for fruitful discussions.


\newpage
\begin{thebibliography}{999}
%
%

\bibitem{SW.86} B.D. Serot and J.D. Walecka,
    Adv. Nucl. Phys. \textbf{16} (1986) 1.

\bibitem{Rein.89} P.G. Reinhard, Rep. Prog. Phys. \textbf{52} (1989) 439.

\bibitem{Ser.92} B.D. Serot, Rep. Prog. Phys. \textbf{55} (1992) 1855.

\bibitem{Ring.96} P. Ring, Prog. Part. Nucl. Phys. \textbf{37} (1996) 193.

\bibitem{GRT.90}Y.K. Gambhir, P. Ring, and A. Thimet, Ann. Phys. (N.Y.) 
                \textbf{198} (1990) 132.

\bibitem{SNR.93} M.M. Sharma, M.A. Nagarajan and P. Ring, Phys. Lett.
B \textbf{312} (1993) 377.

\bibitem{Lala.97} G.A. Lalazissis, J. K\"onig, and P. Ring, Phys. Rev.
C \textbf{55} (1997) 540.

\bibitem{SLR.93} M.M. Sharma, G.A. Lalazissis, and P. Ring,
    Phys. Lett. B \textbf{317} (1993) 9.
 
\bibitem{SLK.94}M.M. Sharma, G.A. Lalazissis, J. K\"onig,
    and P. Ring, Phys. Rev. Lett. \textbf{74} (1994) 3744.

\bibitem{Tajima.93} N. Tajima, P. Bonche, H. Flocard, P.-H. Heenen and
M.S. Weiss, Nucl. Phys. {\bf A551} (1993) 434.

\bibitem{RF.95} P.-G. Reinhard and H. Flocard, Nucl. Phys. 
\textbf{A584} (1995) 467.

\bibitem{BB.77} J. Boguta and A.R. Bodmer, Nucl. Phys. 
\textbf{A292} (1977) 413.

\bibitem{Sharma.00} M.M. Sharma, A.R. Farhan and S. Mythili, 
Phys. Rev. C \textbf{61} (2000) 054306.

\bibitem{Suga.94} Y. Sugahara and H. Toki, Nucl. Phys. {\bf A579} (1994) 557.

\bibitem{Toki.98} H. Toki, H. Shen, K. Suniyoshi, H. Sugahara and I. Tanihata,
 J. Phys. G: Nucl. Part. Phys. \textbf{24} (1998) 1479.

\bibitem{Nikolaus.92} B.A. Nikolaus, T. Hoch, and D.G. Madland,
Phys. Rev. C \textbf{46} (1992) 1757.

\bibitem{Rusnak.97} J.J. Rusnak and R.J. Furnstahl, Nucl. Phys. 
\textbf{A626} (1997) 495.

\bibitem{Buerv.02} T. B\"urvenich, D.G. Madland, J.A. Maruhn, and
  P.G. Reinhard, Phys. Rev. C \textbf{65} (2002) 044308.

\bibitem{Brock.92} R. Brockmann and H. Toki, Phys. Rev. 
                  Lett. \textbf{68} (1992) 3408.

\bibitem{Typel.99} S. Typel and H.H. Wolter, Nucl. Phys. 
\textbf{A656} (1999) 331.


\bibitem{Typel.05} S. Typel, Phys. Rev. C \textbf{71} (2005) 064301.

\bibitem{Niks.02} T. Niksic, D. Vretenar, P. Finelli and P. Ring,
                 Phys. Rev. C \textbf{66} (2002) 024306.

\bibitem{Long.04} W. Long, J. Meng, N. Van Giai and S.G. Zhou,
                 Phys. Rev. C \textbf{69} (2004) 034319.

\bibitem{Savush.97} L.N. Savushkin et al, Phys. Rev. C \textbf{55} (1997) 167.

\bibitem{Furn.97} R.J. Furnstahl, B.D. Serot, and H.B. Tang, 
                 Nucl. Phys. A \textbf{615} (1997) 441.

\bibitem{Serot.04} B.D. Serot, in {\it Lecture Notes in Physics} Vol. 641, 
  (Springer Verlag, 2004) p. 31.

\bibitem{Todd.05} B.G. Todd-Rutel, J. Pieckarewicz, Phys. Rev. 
                  Lett. \textbf{95} (2005) 122501.

\bibitem{Audi.95} G. Audi and A.H. Wapstra, Nucl. Phys. {\bf A595} (1995) 409.

\bibitem{Berger.84} J.F. Berger, M. Girod, and D. Gogny, Nucl. Phys. 
{\bf A428} (1984) 32.

\bibitem{Otten.89} E.W. Otten, in {\it Treatise on Heavy-Ion Science},
edited by D.A. Bromley (Plenum, New York, 1989) Vol 7, p. 517.

\bibitem{Jurado.07} B. Jurado et al., Phys. Lett. B \textbf{649} (2007) 43.

\bibitem{Sikler.05} G. Sikler et al., Nucl. Phys. {\bf A763} (2005) 45.

\bibitem{Hager.06} U. Hager et al., Phys. Rev. Lett. \textbf{96} (2006) 122501.

\bibitem{Sharma.07} M.M. Sharma, (unpublished, 2008)

\bibitem{Sto.94} M.V. Stoitsov, P. Ring and M.M. Sharma, Phys. Rev.
                 {\bf C50} (1994) 1445.

\bibitem{Young04} D.H. Youngblood, H.L.Clark, Y.W. Lui,
          Phys. Rev. C69 (2004) 034315.

\bibitem{Sharma.88} M.M. Sharma et al., Phys. Rev {\bf C38} (1988) 2562.

\bibitem{Klaehn.06} T. Kl\"ahn et al., Phys. Rev {\bf C74} (2006) 035802.

\bibitem{Oezel.06} F. \"Ozel, Nature {\bf 441} (2006) 1115.

\bibitem{FP.81} B. Friedman and V.R. Pandharipande, 
       Nucl. Phys. {\bf A361} (1981) 502.


\end{thebibliography}
\end{document}